\RequirePackage{lineno}
\documentclass[aps,twocolumn,showpacs,byrevtex]{revtex4-1}
\usepackage{times}
\usepackage{amsmath}
\usepackage{graphicx}
\usepackage{color}
\newcommand{\y}{Y(4360)}
\newcommand{\z}{Z_c(3900)}
\newcommand{\x}{X(3823)}
\newcommand{\pp}{\pi^+\pi^-}

\newcommand{\LL}{\ell^+\ell^-}
\newcommand{\EE}{e^+e^-}
\newcommand{\MM}{\mu^+\mu^-}
\newcommand{\GG}{\gamma\gamma}
\newcommand{\etap}{\eta^\prime}
\newcommand{\psip}{\psi^\prime}

\newcommand{\jpsi}{J/\psi}
\newcommand{\piz}{\pi^0}
\newcommand{\chico}{\chi_{c1}}
\newcommand{\chict}{\chi_{c2}}

\newcommand{\psifiv}{\psi(4415)}

\newcommand{\ppjpsi}{\pi^+\pi^-J/\psi}

\def\Journal#1#2#3#4{{#1} {\bf #2}, #3 (#4)}

\def\PRL{Phys. Rev. Lett.}
\def\PRD{Phys. Rev. D}

\def\EPJC{Eur. Phys. J. C}

\begin{document}

\title{\boldmath Observation of the $\psi(1^3D_2)$ state in $e^+e^-\to\pi^+\pi^-\gamma\chi_{c1}$ at BESIII}

\author{
M.~Ablikim$^{1}$, M.~N.~Achasov$^{9,a}$, X.~C.~Ai$^{1}$, O.~Albayrak$^{5}$, M.~Albrecht$^{4}$, D.~J.~Ambrose$^{44}$, A.~Amoroso$^{48A,48C}$, F.~F.~An$^{1}$, Q.~An$^{45}$, J.~Z.~Bai$^{1}$
% ...
\\
\vspace{0.2cm}
(BESIII Collaboration)\\
\vspace{0.2cm} {\it
$^{1}$ Institute of High Energy Physics, Beijing 100049, People's Republic of China\\
% ...
$^{4}$ Bochum Ruhr-University, D-44780 Bochum, Germany\\
$^{5}$ Carnegie Mellon University, Pittsburgh, Pennsylvania 15213, USA\\
% ...
$^{9}$ G.I. Budker Institute of Nuclear Physics SB RAS (BINP), Novosibirsk 630090, Russia\\
% ...
$^{44}$ University of Rochester, Rochester, New York 14627, USA\\
$^{45}$ University of Science and Technology of China, Hefei 230026, People's Republic of China\\
% ...
$^{48}$ (A)University of Turin, I-10125, Turin, Italy; (B)University of Eastern Piedmont, I-15121, Alessandria, Italy; (C)INFN, I-10125, Turin, Italy\\
% ...
$^{a}$ Also at the Novosibirsk State University, Novosibirsk, 630090, Russia\\
% ...
}
}

\vspace{0.4cm}

\date{\today}

\begin{abstract}

We report the observation of the $X(3823)$ in the process 
$e^+e^-\to \pi^+\pi^-X(3823) \to \pi^+\pi^-\gamma\chi_{c1}$
with a statistical significance of $6.2\sigma$, in data 
samples at center-of-mass energies $\sqrt{s}=$4.230, 4.260, 4.360, 4.420 and 4.600~GeV 
collected with the BESIII detector at the BEPCII 
electron positron collider. The measured mass of the $X(3823)$ is
$(3821.7\pm 1.3\pm 0.7)$~MeV/$c^2$, where the first error is
statistical and the second systematic, and the width is less than
$16$~MeV at the 90\% confidence level. The products
of the Born cross sections for $e^+e^-\to \pi^+\pi^-X(3823)$ 
and the branching ratio $\mathcal{B}[X(3823)\to \gamma\chi_{c1,c2}]$ are 
also measured. These measurements are in good agreement with 
the assignment of the $X(3823)$ as the $\psi(1^3D_2)$ charmonium
state.

\end{abstract}


\maketitle

Since its discovery, charmonium - meson particles which contain a charm
and an anti-charm quark - has been an excellent tool for probing
Quantum Chromodynamics (QCD), the fundamental theory that describes
the strong interactions between quarks and gluons, in the 
non-perturbative (low-energy, long-distance effects) regime, and 
remains of high interest both experimentally and theoretically. 
All of the charmonium states with masses that are below 
the open-charm threshold have been firmly established~\cite{potential, pdg};
open-charm refers to mesons containing a charm quark (antiquark) and either 
an up or down antiquark (quark), such as $D$ or $\bar{D}$. 
However, the observation of the spectrum that are above the open-charm 
threshold remains unsettled. During the past decade, many
new charmoniumlike states were discovered, such as the
$X(3872)$~\cite{x3872}, the $Y(4260)$~\cite{y4260,bellezc} 
and the $\z$~\cite{zc3900,bellezc,Xiao:2013iha}. These states provide
strong evidence for the existence of exotic hadron
states~\cite{epjc-review}. Although charged charmoniumlike 
states like the $Z_c(3900)$ provide convincing evidence for 
the existence of multi-quark states~\cite{four-quark}, 
it is more difficult to distinguish neutral candidate exotic states 
from conventional charmonium. Moreover, the study of 
transitions between charmonium(like) states, such as the 
$Y(4260)\to\gamma X(3872)$~\cite{bes3-x3872}, is an important
approach to probe their nature, and the connections between them.
Thus, a more complete understanding of the charmonium(like) 
spectroscopy and their relations is necessary and timely.

The lightest charmonium state above the $D\bar{D}$ threshold is
the $\psi(3770)$~\cite{pdg}, which is currently identified as the $1^3D_1$
state~\cite{potential}, the $J=1$ member of the $D$-wave spin-triplet
charmonium states. Until now there have been no definitive observations of
its two $D$-wave spin-triplet partner states, i.e., the $1^3D_2$
and $1^3D_3$. Phenomenological models predict that the $1^3D_2$
charmonium state has large decay widths to $\gamma\chico$ and
$\gamma\chict$~\cite{3d2-decay}. In 1994, the E705 experiment reported
a candidate for the $1^3D_2$ state with a mass of $3836\pm
13$~MeV/$c^2$ and a statistical significance of
$2.8\sigma$~\cite{E705}. Recently, the Belle Collaboration
reported evidence for a narrow resonance $\x\to \gamma\chico$
in $B$ meson decays with $3.8\sigma$ significance and mass 
$3823.1\pm 1.8 {\rm (stat)}\pm 0.7 {\rm (syst)}$~MeV/$c^2$, 
and suggested that this is a good candidate for the $1^3D_2$ 
charmonium state~\cite{belle-3d2}. In the following, we denote 
the $1^3D_2$ state as $\psi_2$ and the $\psi(3686)$ [$\psi(2S)$] state as $\psip$.

In this Letter, we report a search for the production of the $\psi_2$ state
via the process $\EE\to \pp X$, using 4.67~fb$^{-1}$ data collected with the BESIII
detector operating at the BEPCII storage ring~\cite{bepc2} at
center-of-mass (CM) energies that range from $\sqrt{s}=4.19$ to
4.60~GeV~\cite{lum}. The $\psi_2$ candidates are
reconstructed in their $\gamma\chico$ and $\gamma\chict$ decay modes,
with $\chi_{c1,c2}\to \gamma J/\psi$ and $J/\psi\to \LL$ ($\ell=e$
or $\mu$).
%
%
A {\sc geant4}-based~\cite{geant4} Monte Carlo~(MC) simulation 
software package 
is used to optimize event selection
criteria, determine the detection efficiency, and estimate the
backgrounds. For the signal process, we generate 40,000 $\EE\to \pp\x$ 
events at each CM energy indicated above, 
using an {\sc evtgen}~\cite{evtgen} phase space model, with $\x\to \gamma\chi_{c1,c2}$.
Initial state radiation (ISR) is simulated with {\sc kkmc}~\cite{kkmc}, where
the Born cross section of $\EE\to \pp\x$ between 4.1 and 4.6~GeV
is assumed to follow the $\EE\to \pp\psip$
lineshape~\cite{y4360}. The maximum ISR photon energy is set to correspond
to the 4.1~GeV/$c^2$ production threshold of the $\pp\x$ system.
Final-State-Radiation is handled with {\sc
photos}~\cite{photos}.

Events with four charged tracks with zero net charge are
selected as described in Ref.~\cite{zc3900}. 
%
Showers identified as photon candidates must satisfy fiducial and
shower quality as well as timing requirements as described in Ref.~\cite{bes3-etajpsi}.
At least two good photon candidates in each event are required.
To improve the momentum and energy resolution and to reduce
the background, the event is subjected to a four-constraint~(4C)
kinematic fit to the hypothesis $\EE\to \pi^+ \pi^-\gamma\gamma
\ell^+ \ell^-$, that constrains the total four-momentum of the
detected particles to the initial four-momentum of the
colliding beams. The $\chi^2$ of the kinematic fit is required to
be less than 80 (with an efficiency of about 95\% for
signal events). For multi-photon events, the two photons
returning the smallest $\chi^2$ from the 4C fit are assigned to 
be the radiative photons. 

To reject radiative Bhabha and radiative dimuon
($\gamma\EE/\gamma\MM$) backgrounds associated with
photon conversion, the cosine of the opening angle of the pion-pair
candidates is required to be less than 0.98. This restriction
removes almost all Bhabha and dimuon background events, with an
efficiency loss that is less than 1\% for signal events. The background from
$\EE\to \eta\jpsi$ with $\eta\to \pp\pi^0/\gamma\pp$ is effectively rejected
by the invariant mass requirement $M(\GG\pp)>0.57~{\rm GeV}/c^2$. 
MC simulation shows that this requirement removes less than 1\% 
of the signal events. In order to remove
possible backgrounds from $\EE\to\gamma_{\rm ISR}\psip\to
\gamma_{\rm ISR}\ppjpsi$, accompanied with a fake photon or a second
ISR photon, $\EE\to\eta\psip$ with $\eta\to\GG$, and $\EE\to\GG\psip$, 
the invariant mass of $\ppjpsi$ is required to
satisfy $|M(\ppjpsi)-m(\psip)|>6$~MeV/$c^2$~\cite{Mppjpsi}. 
The signal efficiency
for the $\psip$ mass window veto is $85\%$ at $\sqrt{s}=4.420$~GeV
and $\ge 99\%$ at other energies.

After imposing the above requirements, there are clear $\jpsi$
peaks in the $M(\LL)$ invariant mass distributions for the data. The
$\jpsi$ mass window is defined as $3.08<M(\LL)<3.13$~GeV/$c^2$.
The mass resolution is determined to be 9~MeV/$c^2$ by MC simulation.
In order to evaluate non-$\jpsi$ backgrounds, we define 
$\jpsi$ mass sidebands as $3.01<M(\LL)<3.06$~GeV/$c^2$ or
$3.15<M(\LL)<3.20$~GeV/$c^2$, which are twice as wide as the signal
region. The combination of the higher energy photon ($\gamma_H$) with
the $\jpsi$ candidate is used to reconstruct $\chi_{c1,c2}$ signals, while
the lower one is assumed to originate from the $\x$ decay. 
We define the invariant mass range
$3.490<M(\gamma_H\jpsi)<3.530$~GeV/$c^2$
as the $\chico$ signal region, and
$3.536<M(\gamma_H\jpsi)<3.576$~GeV/$c^2$ 
as the $\chict$ signal region [$M(\gamma_H\jpsi)=M(\gamma_H\LL)-M(\LL)+m(\jpsi)$].

\begin{figure}
\begin{center}
\includegraphics[height=2.9in]{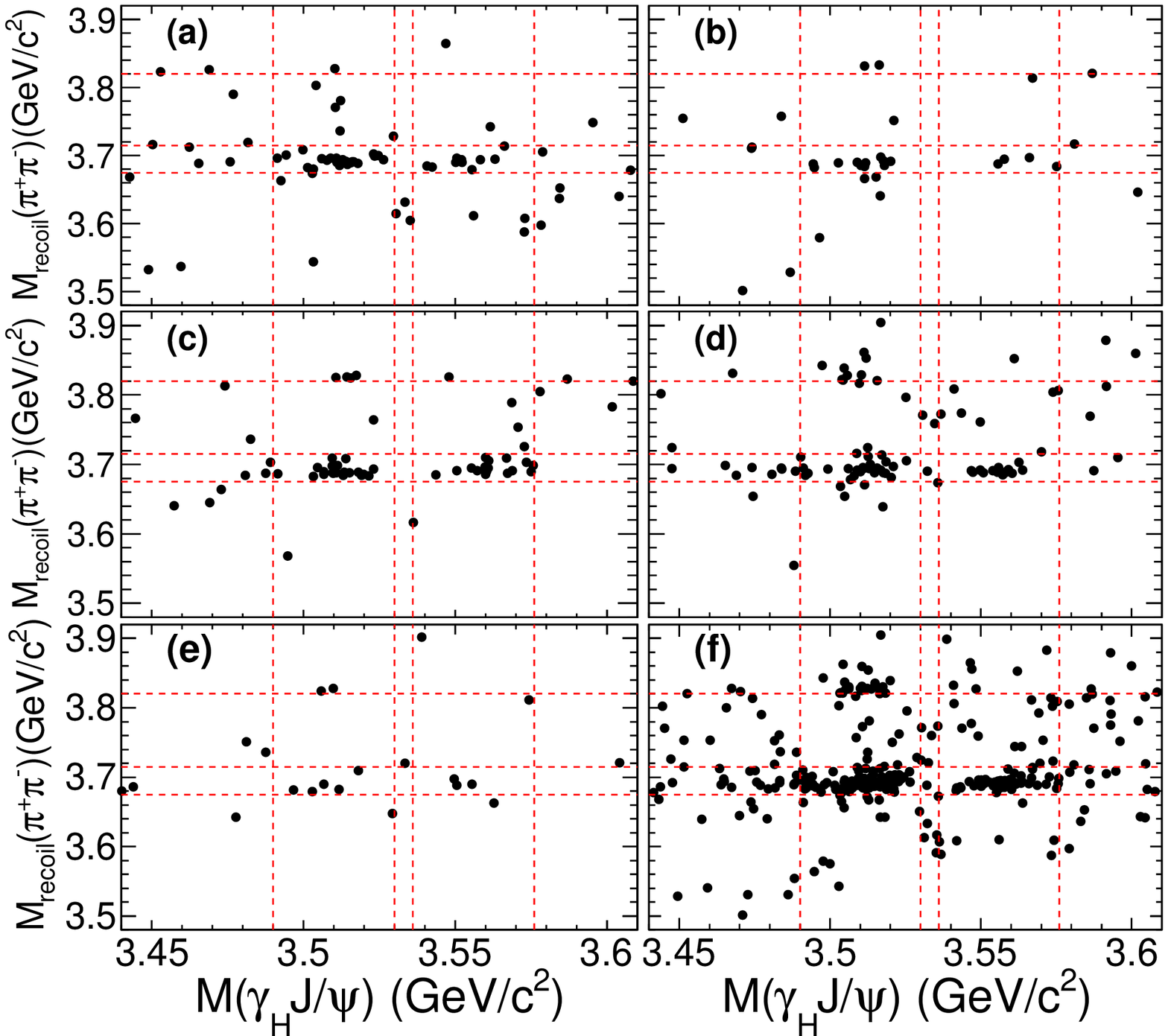}
\caption{Scatter plots of $M_{\rm recoil}(\pp)$ vs. $M(\gamma_H\jpsi)$ 
at (a) $\sqrt{s}=$4.230, (b) 4.260, (c) 4.360, (d) 4.420, and (e) 4.600~GeV. 
The sum of all the data sets is shown in (f). In each plot, the 
vertical dashed red lines represent $\chi_{c1}$ (left two lines) and 
$\chi_{c2}$ (right two lines) signal regions, and the horizontal lines represent the
$\psip$ mass range (bottom two lines) and 3.82~GeV (top line), respectively.}
\label{recoil}
\end{center}
\end{figure}

To investigate the possible existence of resonances that may decay to
$\gamma\chi_{c1,c2}$, we examine two-dimensional scatter plots of
$M_{\rm recoil}(\pp)$ versus $M(\gamma_H\jpsi)$. Here, $M_{\rm
recoil}(\pp)=\sqrt{(P_{\EE}-P_{\pi^+}-P_{\pi^-})^2}$ is the recoil
mass of the $\pp$ pair, where $P_{\EE}$ and $P_{\pi^\pm}$ are the 4-momenta 
of the initial $\EE$ system and the $\pi^\pm$, respectively. 
For this, we use the $\pp$ momenta before the 4C fit correction
because of the good resolution for low momentum
pion tracks, as observed from MC simulation. Figure~\ref{recoil}
shows $M_{\rm recoil}(\pp)$ versus
$M(\gamma_H\jpsi)$ for data at different energies, where
$\EE\to \pp\psip\to \pp\gamma\chi_{c1,c2}$ signals are evident in
almost all data sets. In addition, event accumulations 
near $M_{\rm recoil}(\pp)\simeq 3.82$~GeV$/c^2$ are evident 
in the $\chico$ signal regions of the $\sqrt{s}=4.36$ and 4.42~GeV data sets. 
A scatter plot of all the data sets combined is shown in Fig.~\ref{recoil} (f), 
where there is a distinct cluster of events near $3.82$~GeV/$c^2$ (denoted
hereafter as the $\x$) in the $\chico$ signal region.

The remaining backgrounds mainly come from
$\EE\to(\etap/\gamma\omega)\jpsi$, with $(\etap/\omega)\to
\GG\pp/\gamma\pp$, and $\pp\pp(\piz/\GG)$. 
The $\EE\to(\etap/\gamma\omega)\jpsi$
backgrounds can be measured and simulated 
using the same data sets. The $\EE\to \pp\pp(\piz/\GG)$ mode
can be evaluated with the $\jpsi$ mass sideband data.
All these backgrounds are found to be small, and they
produce flat contributions to the $M_{\rm recoil}(\pp)$ mass
distribution. There also might be $\EE\to\pp\psip$ events
with $\psip\to\eta\jpsi$ and $\pi^0\pi^0\jpsi$, but such kind of
events would not affect the $\psip$ mass in the
$M_{\rm recoil}(\pp)$ distribution.

An unbinned maximum likelihood fit to the $M_{\rm recoil}(\pp)$
invariant mass distribution is performed to extract the $\x$
signal parameters. The signal shapes are represented by MC-simulated $\psip$
and $\x$ (with input mass of 3.823~GeV/$c^2$ and a zero width)
histograms, convolved with Gaussian functions with mean and width
parameters left free in the fit to
account for the mass and resolution difference between data and MC
simulation, respectively. The background is parameterized as a linear
function, as indicated by the $\jpsi$ mass sideband data.
The $\psip$ signal is used to
calibrate the absolute mass scale and the resolution difference
between data and simulation, which is expected to be similar for the 
$\x$ and $\psip$. A simultaneous fit with a common $\x$ mass 
is applied to the data sets with independent signal yields at
$\sqrt{s}=4.230$, 4.260, 4.360, 4.420 and 4.600~GeV (data
sets with small luminosities are merged to nearby
data sets with larger luminosities), 
for the $\gamma\chico$ and $\gamma\chict$ modes, respectively.

\begin{figure}
\begin{center}
\includegraphics[height=1.2in]{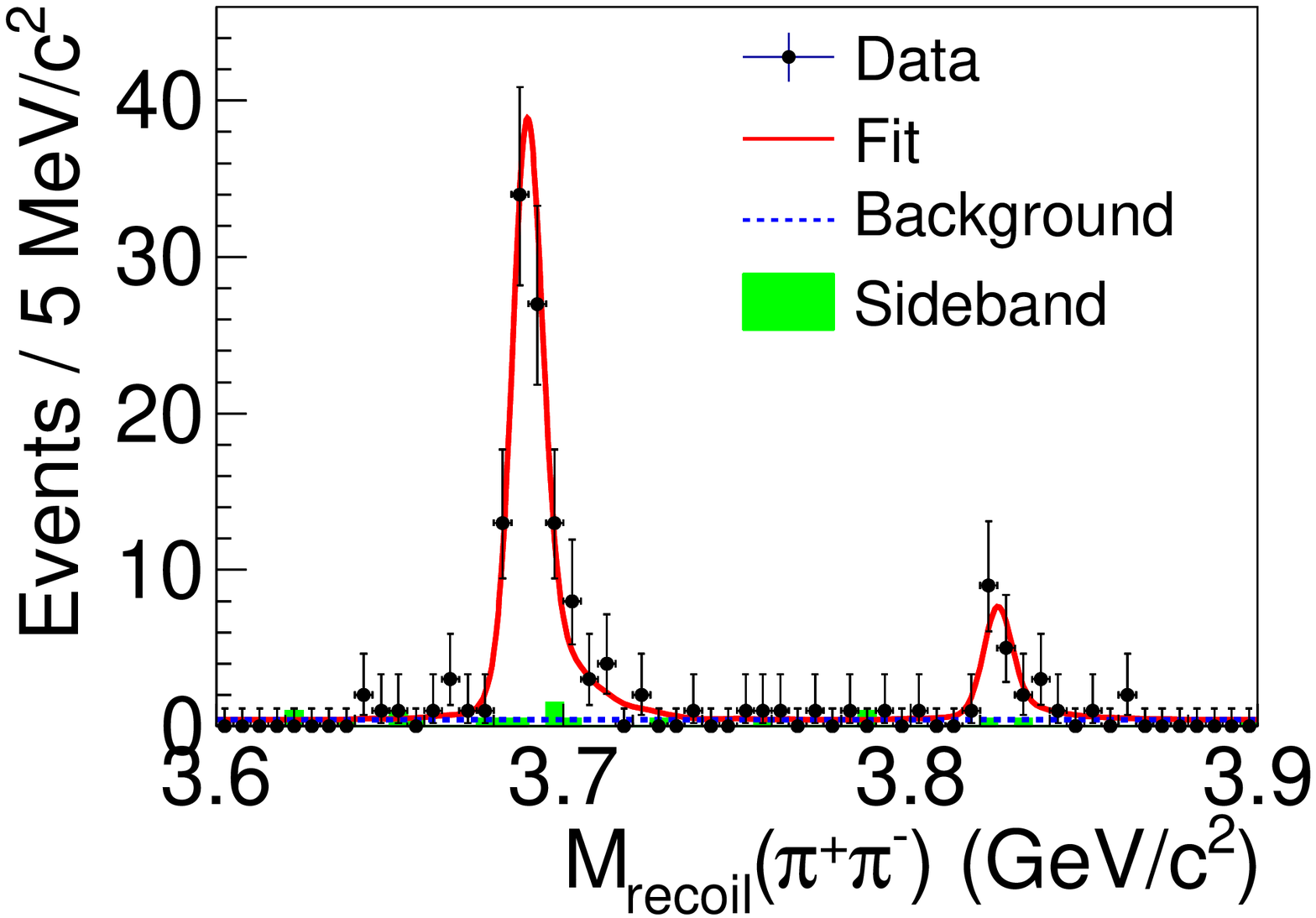}
\includegraphics[height=1.2in]{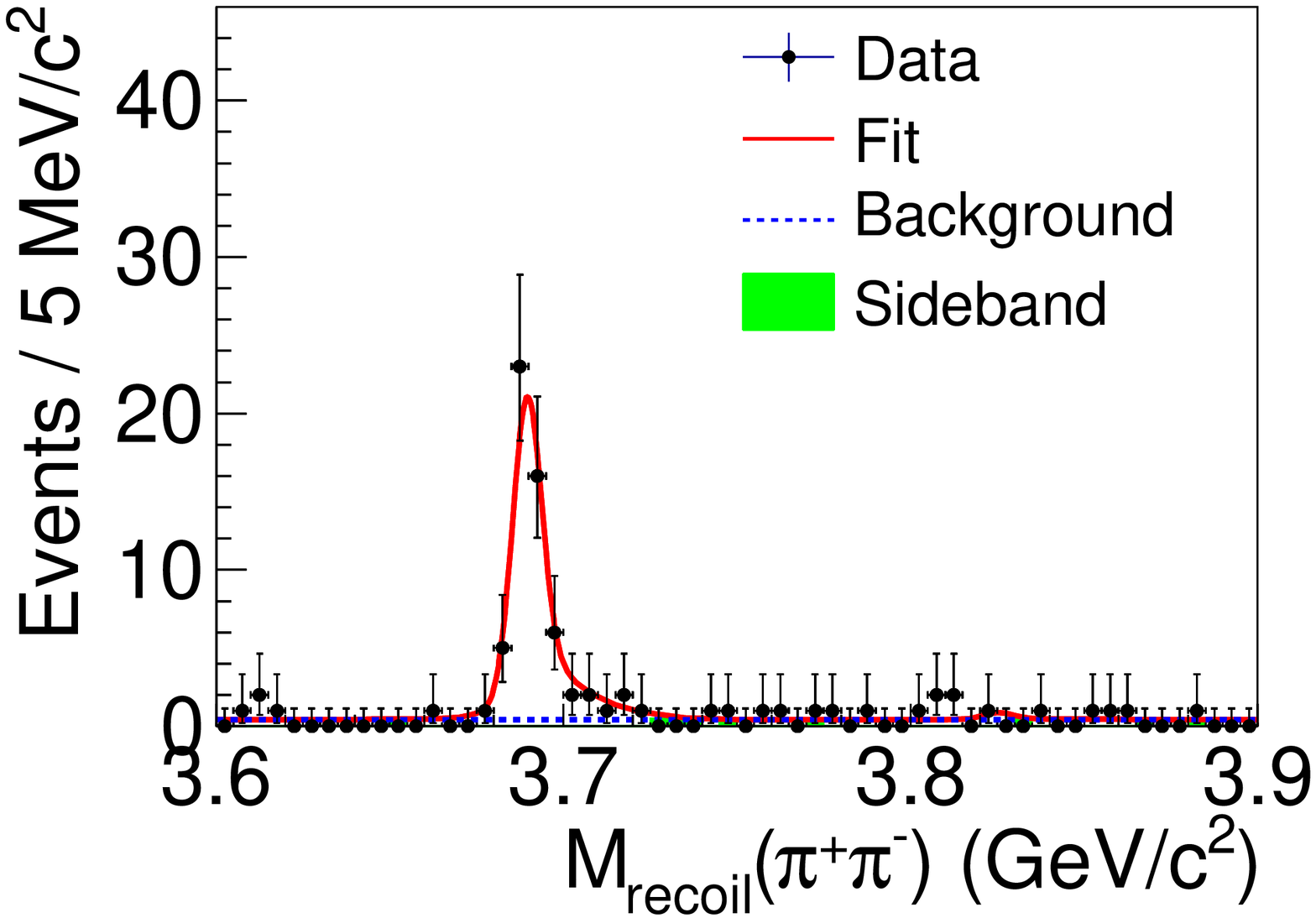}
\caption{Simultaneous fit to the $M_{\rm recoil}(\pp)$
distribution of $\gamma\chico$ events (left) and
$\gamma\chict$ events (right), respectively. 
Dots with error bars are data, red solid curves are total fit, dashed
blue curves are background, and the green shaded histograms are
$\jpsi$ mass sideband events.} \label{X-fit}
\end{center}
\end{figure}

Figure~\ref{X-fit} shows the fit results, which return
$M[\x]=M[\x]_{\rm input}+\mu_{\x}-\mu_{\psip} = 3821.7\pm
1.3$~MeV/$c^2$ for the $\gamma\chico$ mode, where $M[\x]_{\rm input}$
is the input $\x$ mass in MC simulation, $\mu_{\x}=1.9\pm1.3$~MeV/$c^2$ and
$\mu_{\psip}=3.2\pm0.6$~MeV/$c^2$ are the mass shift values for $\x$ and $\psip$
histograms from the fit. The fit yields $19\pm 5$ $\x$ signal
events in the $\gamma\chico$ mode. The statistical significance of
the $\x$ signal in the $\gamma\chico$ mode
is estimated to be $6.2\sigma$ by
comparing the difference between the log-likelihood value
($\Delta(\ln\mathcal{L})=27.5$) with or without $\x$ signal
in the fit, and taking the change of the number of degrees of freedom
($\Delta {\rm ndf}=6$) into account, and its value is found to be larger
than $5.9\sigma$ with various systematic checks.
For the $\gamma\chict$ mode, we
do not observe an $\x$ signal and provide an upper limit
on its production rate (Table~\ref{sec}). The limited statistics
preclude a measurement of the intrinsic width of $\x$. From a fit
using a Breit-Wigner function (with a width parameter that is allowed to float)
convolved with Gaussian resolution, we determine 
$\Gamma[\x]<16$~MeV at the 90\% confidence level (C.L.) 
(including systematic errors).

\begin{figure}
\begin{center}
\includegraphics[height=1.2in]{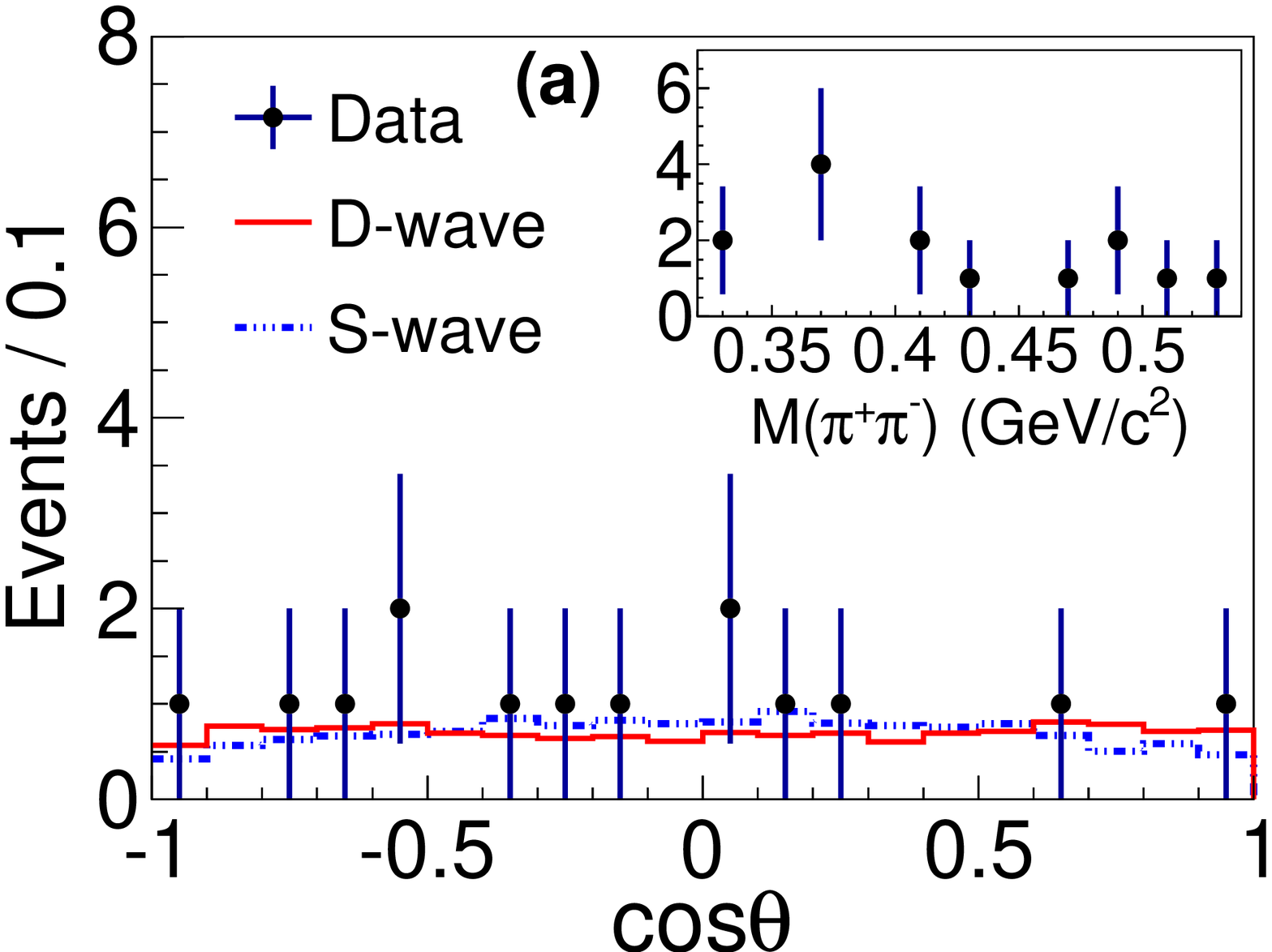}
\includegraphics[height=1.2in]{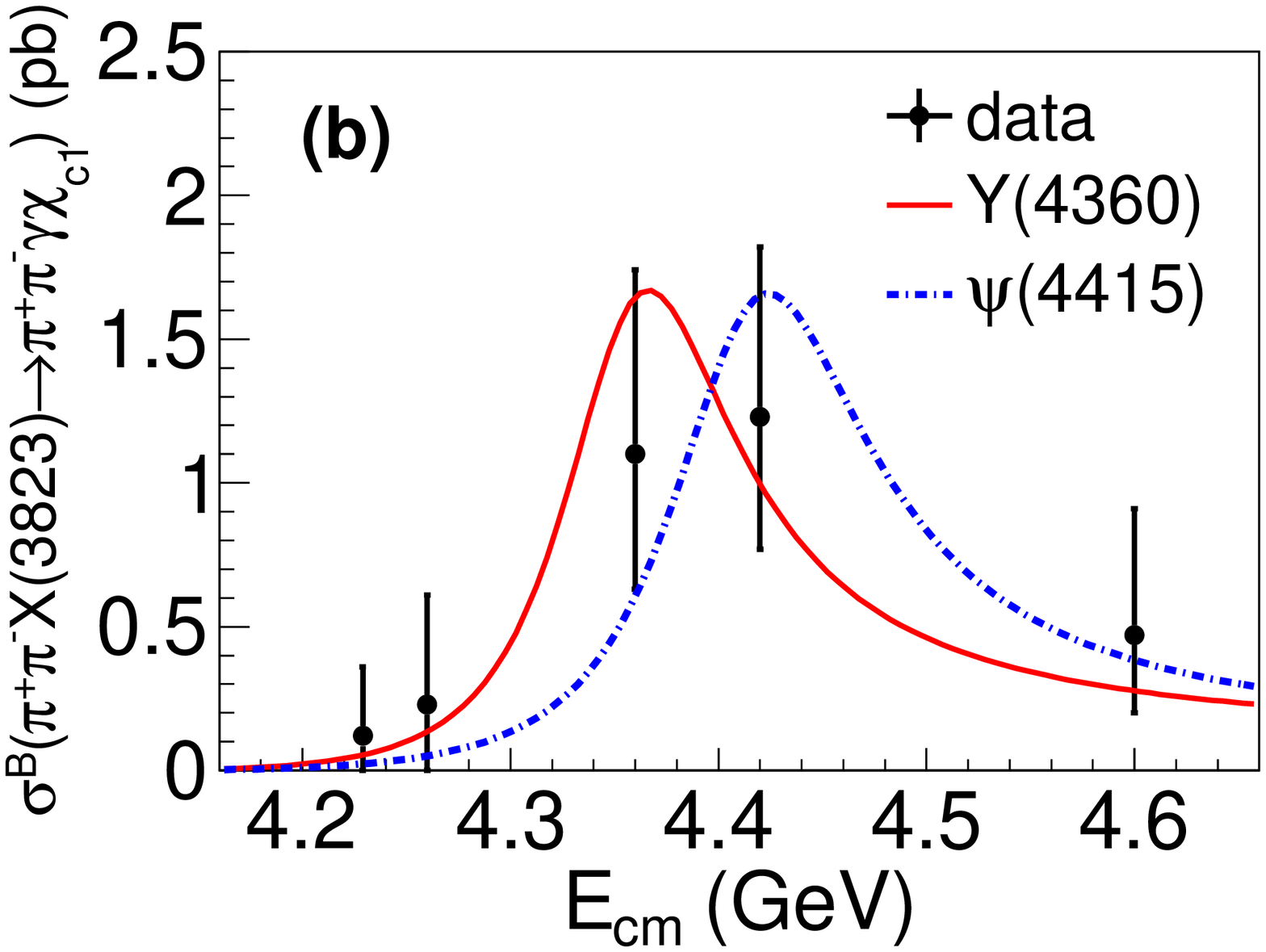}
\caption{
(a) The $\x$ scattering angle distribution for $\x$ signal events, the inset shows
the corresponding $M(\pp)$ invariant mass distribution per 20~MeV/$c^2$ bin; and
(b) fit to the energy-dependent cross section of
$\sigma^B[\EE\to\pp\x]\cdot\mathcal{B}(\x\to\gamma\chico)$ with the 
$Y(4360)$ (red solid curve) and the $\psi(4415)$ (blue dashed curve) lineshapes. 
Dots with error bars are data. The red solid (blue dashed) histogram
in (a) is MC simulation with $D$-wave ($S$-wave).
} 
\label{cos-sec}
\end{center}
\end{figure}

The $\x$ is a candidate for the $\psi_2$ charmonium state
with $J^{PC}=2^{--}$~\cite{belle-3d2}. In the $\EE\to\pp\psi_2$ process, 
the $\pp$ system is very likely to be dominated by $S$-wave.
Thus, a $D$-wave between the $\pp$ system and $\psi_2$ is expected, with
an angular distribution of $1+\cos^2\theta$ for $\psi_2$ in the $\EE$
CM frame. Figure~\ref{cos-sec} (a) shows the angular distribution ($\cos\theta$)
of $\x$ signal events selected by requiring $3.82<M_{\rm recoil}(\pp)<3.83$~GeV/$c^2$.
The inset shows the corresponding $M(\pp)$ invariant mass distribution 
per 20~MeV/$c^2$ bin.
A Kolmogorov~\cite{Kolmogorov} test to the angular distribution
gives the Kolmogorov statistic $D_{14,\rm obs}^{D}=0.217$ for the $D$-wave 
hypothesis and $D_{14,\rm obs}^{S}=0.182$ for the $S$-wave hypotheses. Due to
limited statistics, both hypothesis can be accepted 
($D_{14,\rm obs}^{D},D_{14,\rm obs}^{S}<D_{14,0.1}=0.314$) at the 90\% C.L. 

The product of the Born-order cross section and the branching
ratio of $\x\to\gamma\chi_{c1,c2}$ is calculated using
$\sigma^{B}[\EE\to\pp \x]\cdot \mathcal{B}[\x\to
\gamma\chi_{c1,c2}] = \frac{N^{\rm obs}_{c1,c2}} {\mathcal{L}_{\rm int}
(1+\delta) \frac{1}{|1-\Pi|^2} \epsilon \mathcal{B}_{c1,c2}}$, where
$N^{\rm obs}_{c1,c2}$ is the number of $\x\to\gamma\chi_{c1,c2}$ signal events 
obtained from a fit to the $M_{\rm recoil}(\pp)$ distribution,
$\mathcal{L}_{\rm int}$ is the integrated luminosity, $\epsilon$ is the detection
efficiency, $\mathcal{B}_{c1,c2}$ is the branching fraction of
$\chi_{c1,c2}\to \gamma\jpsi\to \gamma \LL$ and ($1+\delta$) is
the radiative correction factor, which depends on the lineshape
of $\EE\to \pp\x$.
Since we observe large cross sections at
$\sqrt{s}=4.360$ and 4.420~GeV, we assume the $\EE\to \pp\x$ cross
section follows that of $\EE\to \pp\psip$ over the full energy
range of interest and use the $\EE\to \pp\psip$ lineshape from
published results~\cite{y4360} as input in the calculation of
the efficiency and radiative correction factor. The vacuum
polarization factor $\frac{1}{|1-\Pi|^2}$ is calculated from QED
with 0.5\% uncertainty~\cite{VP}. The results of these
measurements for the data sets with large luminosities at $\sqrt{s}=4.230$, 4.260,
4.360, 4.420 and 4.600~GeV are listed in Table~\ref{sec}. Since
at each single energy data the $\x$ signal is not very significant,
upper limits for production cross sections at the 90\%
C.L. based on the Bayesian method are given
[systematic effects are included by convolving the $\x$ signal
events yield ($n^{\rm yield}$) dependent likelihood curves 
with a Gaussian with mean value zero and standard deviation 
$n^{\rm yield}\cdot\sigma_{\rm sys}$,
where $\sigma_\mathrm{sys}$
is the systematic uncertainty of the efficiencies]. 
The corresponding production ratio of $\mathcal{R}_{\psip} =
\frac{\sigma^B[\EE\to\pp\x]\cdot\mathcal{B}[\x\to\gamma\chico]}
{\sigma^B[\EE\to\pp\psip]\cdot\mathcal{B}[\psip\to\gamma\chico]}$
is also calculated at $\sqrt{s}=4.360$ and 4.420~GeV.

\begin{table*}
\begin{center}
\caption{Number of observed events ($N^{\rm obs}$), integrated
luminosities ($\mathcal{L}$)~\cite{lum}, detection
efficiency ($\epsilon$) for the $\x\to\gamma\chico$ mode, radiative
correction factor ($1+\delta$), vacuum polarization factor
($\frac{1}{|1-\Pi|^2}$), measured Born cross section
$\sigma^{B}(\EE\to \pp\x)$ times $\mathcal{B}_1(\x\to
\gamma\chico)$ ($\sigma^B_{X}\cdot \mathcal{B}_1$) and
$\mathcal{B}_2(\x\to \gamma\chict)$ ($\sigma^B_{X}\cdot
\mathcal{B}_2$), and measured Born cross section $\sigma^B(\EE\to
\pp\psip)$ ($\sigma^B_{\psip}$) at different energies. 
Other data sets with lower luminosity are not listed. 
The numbers in the brackets correspond to the upper limit 
measurements at the 90\% C.L. The relative ratio $\mathcal{R}_{\psip} =
\frac{\sigma^B[\EE\to\pp\x]\mathcal{B}(\x\to\gamma\chico)}
{\sigma^B[\EE\to\pp\psip]\mathcal{B}(\psip\to\gamma\chico)}$ is
also calculated. The first errors are statistical, and the second systematic.} \label{sec}
\begin{tabular}{cccccccccc}
  \hline\hline
  $\sqrt{s}$~(GeV) & $\mathcal{L}$~(pb$^{-1}$) & $N^{\rm obs}$ & $\epsilon$ & $1+\delta$ & $\frac{1}{|1-\Pi|^2}$ & $\sigma^B_X\cdot\mathcal{B}_1$~(pb) & $\sigma^B_X\cdot\mathcal{B}_2$~(pb) & $\sigma^B_{\psip}$ (pb) & $\mathcal{R}_{\psip}$ \\
  \hline
  4.230 & 1092 & $0.7^{+1.4}_{-0.7}$ $(<3.8)$ & 0.168 & 0.755 & 1.056 & $0.12^{+0.24}_{-0.12}\pm 0.02$ $(<0.64)$ & - & $34.1\pm8.1\pm4.7$ &- \\
  4.260 & 826  & $1.1^{+1.8}_{-1.2}$ $(<4.6)$ & 0.178 & 0.751 & 1.054 & $0.23^{+0.38}_{-0.24}\pm 0.04$ $(<0.98)$ & - & $25.9\pm8.1\pm3.6$ & - \\
  4.360 & 540  & $3.9^{+2.3}_{-1.7}$ $(<8.2)$ & 0.196 & 0.795 & 1.051 & $1.10^{+0.64}_{-0.47}\pm 0.15$ $(<2.27)$ & $(<1.92)$ & $58.6\pm14.2\pm8.1$ & $0.20^{+0.13}_{-0.10} $ \\
  4.420 & 1074 & $7.5^{+3.6}_{-2.8}$ $(<13.4)$ & 0.145 & 0.967 & 1.053 & $1.23^{+0.59}_{-0.46}\pm 0.17$ $(<2.19)$ & $(<0.54)$ & $33.4\pm7.8\pm4.6$ & $0.39^{+0.21}_{-0.17}$ \\
  4.600 & 567  & $1.9^{+1.8}_{-1.1}$ $(<5.4)$ & 0.157 & 1.075 & 1.055 & $0.47^{+0.44}_{-0.27}\pm 0.07$ $(<1.32)$ & - & $10.4^{+6.4}_{-4.7}\pm1.5$ & - \\
  \hline\hline
\end{tabular}
\end{center}
\end{table*}

We fit the energy-dependent cross sections of $\EE\to\pp\x$ with
the $\y$ shape or the $\psifiv$ shape with their resonance
parameters fixed to the PDG values~\cite{pdg}.
Figure~\ref{cos-sec} (b) shows the fit results, which give
$D^{\rm H1}_{5,{\rm obs}}=0.151$ for the $\y$ hypothesis (H1) and $D^{\rm
H2}_{5,{\rm obs}}=0.169$ for the $\psifiv$ hypothesis (H2), based
on the Kolmogorov test. Thus, we
accept both the $\y$ and the $\psifiv$ hypotheses ($D^{\rm
H1}_{5,{\rm obs}}, D^{\rm H2}_{5,{\rm obs}}<D_{5,0.1}=0.509$) at
the 90\% C.L.



The systematic uncertainties in the $\x$ mass measurement include
those from the absolute mass scale, resolution, the
parameterization of the $\x$ signal, and the background shape. Since we
use the $\psip$ signal to calibrate the fit, we conservatively take
the uncertainty of 0.6~MeV/$c^2$ in the calibration procedure as the
systematic uncertainty due to the mass scale. The resolution
difference between the data and MC simulation is also estimated by the
$\psip$ signal. Varying the resolution parameter by $\pm 1\sigma$,
the mass difference in the fit is 0.2~MeV/$c^2$,
which is taken as the systematic uncertainty from resolution. In the $\x$
mass fit, a MC-simulated histogram with the width of $\x$ set to zero is used to
parameterize the signal shape. We replace this histogram with a
simulated $\x$ resonance with a width of 1.7~MeV~\cite{belle-3d2}
and repeat the fit; the change in the mass for this
fit, 0.2~MeV/$c^2$, is taken as the systematic uncertainty due to the signal
parameterization. Likewise, changes measured with a background
shape from MC-simulated $(\etap/\gamma\omega)\jpsi$ events or a
second-order polynomial indicate a systematic uncertainty
associated with the background shape of 0.2~MeV/$c^2$ in mass. 
Assuming that all the sources are independent, the total systematic
uncertainty is calculated by adding the individual uncertainties in quadrature, 
resulting in 0.7~MeV/$c^2$ for the $\x$ mass measurement. For the $\x$ width, we
measure the upper limits with the above systematic checks, and
report the most conservative one.

The systematic uncertainties in the cross section measurement
mainly come from efficiencies, signal parameterization, background
shape, decay model, radiative correction, and luminosity measurement. 
The luminosity is measured using Bhabha events, with an uncertainty of
1.0\%. The uncertainty in the tracking efficiency for high momenta
leptons is 1.0\% per track.  Pions have momenta that range from
0.1 to 0.6~GeV/$c$, and the
momentum-weighted uncertainty is 1.0\% per
track. In this analysis, the radiative transition photons have
energies from 0.3 to 0.5~GeV. Studies with a sample of
$\jpsi\to \rho\pi$ events show that the uncertainty in the
reconstruction efficiency for photons in this energy range is less
than 1.0\%.

The same sources of signal parameterization and background shape as discussed in
the systematic uncertainty of $\x$ mass measurement would contribute 4.0\%
and 8.8\% differences in $\x$ signal events yields, which are taken as systematic
uncertainties in the cross section measurement.
Since the $\x$ is a candidate for the $\psi_2$ charmonium state,
we try to model the $\EE\to\pp\x$ process with a $D$-wave in the MC
simulation. The efficiency difference between $D$-wave model and
three-body phase space is 3.8\%, which is quoted
as the systematic uncertainty for the decay model. The $\EE\to \pp\x$
lineshape affects the radiative correction factor and detection
efficiency. The radiator function is calculated from QED with
0.5\% precision~\cite{rad}. As discussed above, both $\y$ 
lineshapes~\cite{y4360,y4360babar} and the $\psifiv$
lineshape describe the cross section of $\EE\to\pp\x$ reasonably
well.  We take the difference for $(1+\delta)\cdot\epsilon$ between
$\y$ lineshapes and the $\psifiv$ lineshape as its systematic uncertainty, 
which is 6.5\%.

Since the event topology in this analysis is quite similar to
$\EE\to\gamma\ppjpsi$~\cite{bes3-x3872}, we use the same systematic
uncertainties for the kinematic fit (1.5\%) and the $\jpsi$ mass
window (1.6\%).
The uncertainties on the branching ratios for $\chi_{c1,c2}\to
\gamma\jpsi$ (3.6\%) and $\jpsi\to \LL$ (0.6\%) are taken from
the PDG~\cite{pdg}.
The uncertainty from MC statistics is 0.3\%.
The efficiencies for other selection criteria, the trigger
simulation~\cite{trigger}, the event-start-time determination, and
the final-state-radiation simulation are very high ($>99\%$), and
their systematic uncertainties are estimated to be less than 1\%.

Assuming that all the systematic uncertainty sources are independent,
we add all of them in quadrature.  The total systematic
uncertainty in the cross section measurements is estimated 
to be 13.8\%.

In summary, we observe a narrow resonance, $\x$, through the
process $\EE\to\pp\x$ with a statistical significance of $6.2\sigma$.
The measured mass of the $\x$ is $(3821.7\pm 1.3\pm 0.7)~{\rm
MeV}/c^2$, where the first error is statistical and the second
systematic, and the width is less than $16$~MeV at the 90\% C.L.
Our measurement agrees well with the values found by
Belle~\cite{belle-3d2}. The production cross
sections of $\sigma^{B}(\EE\to\pp\x)\cdot \mathcal{B}(\x\to
\gamma\chico, \gamma\chict)$ are also measured at
$\sqrt{s}=4.230$, 4.260, 4.360, 4.420, and 4.600~GeV.

The $\x$ resonance is a good candidate for the $\psi(1^3D_2)$
charmonium state. According to potential models~\cite{potential},
the $D$-wave charmonium states are expected to be within a mass
range of 3.82 to 3.85~GeV. Among these, the 
$1^1D_2\to \gamma\chico$ transition is forbidden due to
C-parity conservation, and the amplitude for $1^3D_3\to \gamma\chico$
is expected to be small~\cite{barnes}. The mass of $\psi(1^3D_2)$
is in the $3.810\sim 3.840$~GeV/$c^2$ range that is expected 
for several phenomenological
calculations~\cite{3d2-mass}. In this case, the mass of
$\psi(1^3D_2)$ is above the $D\bar{D}$ threshold but below the
$D\bar{D}^*$ threshold. Since $\psi(1^3D_2)\to D\bar{D}$
violates parity, the $\psi(1^3D_2)$ is expected to be
narrow, in agreement with our observation, and $\psi(1^3D_2)\to\gamma\chico$ 
is expected to be a dominant decay mode~\cite{3d2-mass, ratio}. From our
cross section measurement, the ratio $\frac{\mathcal{B}[\x\to
\gamma\chict]}{\mathcal{B}[\x\to \gamma\chico]}<0.42$ (where systematic
uncertainties cancel) at the 90\% C.L. is obtained, which also agrees with
expectations for the $\psi(1^3D_2)$ state~\cite{ratio}.


The BESIII collaboration thanks the staff of BEPCII and the IHEP
computing center for their strong support. This work is supported in
part by National Key Basic Research Program of China under Contract
No.~2015CB856700; National Natural Science Foundation of China (NSFC)
under Contracts Nos.~11125525, 11235011, 11322544, 11335008, 11425524;
the Chinese Academy of Sciences (CAS) Large-Scale Scientific Facility
Program; the CAS Center for Excellence in Particle Physics (CCEPP); 
Joint Large-Scale Scientific Facility Funds of the NSFC and
CAS under Contracts Nos.~11179007, U1232201, U1332201; CAS under
Contracts Nos.~KJCX2-YW-N29, KJCX2-YW-N45; 100 Talents Program of CAS;
INPAC and Shanghai Key Laboratory for Particle Physics and Cosmology;
German Research Foundation DFG under Contract No.~Collaborative
Research Center CRC-1044; Seventh Framework Programme of the 
European Union under Marie Curie International Incoming Fellowship Grant Agreement No. 627240; 
Istituto Nazionale di Fisica Nucleare,
Italy; Ministry of Development of Turkey under Contract
No.~DPT2006K-120470; Russian Foundation for Basic Research under
Contract No.~14-07-91152; U.S.~Department of Energy under Contracts
Nos.~DE-FG02-04ER41291, DE-FG02-05ER41374, DE-FG02-94ER40823,
DESC0010118; U.S.~National Science Foundation; University of Groningen
(RuG) and the Helmholtzzentrum fuer Schwerionenforschung GmbH (GSI),
Darmstadt; WCU Program of National Research Foundation of Korea under
Contract No.~R32-2008-000-10155-0.



\begin{thebibliography}{**}

\bibitem{potential} E. Eichten, K. Gottfried, T. Kinoshita, K. D. Lane,
and T. M. Yan, Phys. Rev. D {\bf 17}, 3090 (1978); Phys. Rev. D
{\bf 21}, 203 (1980).

\bibitem{pdg} K. A. Olive {\em et al.} (Particle Data Group), Chin. Phys. C {\bf 38}, 090001
(2014).

\bibitem{x3872} S. K. Choi {\em et al.} (Belle Collaboration),
\Journal\PRL{91}{262001}{2003}.

\bibitem{y4260} B. Aubert {\em et al.} (BaBar Collaboration),
Phys. Rev. Lett. {\bf 95}, 142001 (2005).

\bibitem{bellezc} Z.~Q.~Liu {\em et al.} (Belle Collaboration),
\Journal\PRL{110}{252002}{2013}.

\bibitem{zc3900} M. Ablikim {\em et al.} (BESIII Collaboration),
\Journal\PRL{110}{252001}{2013}.

\bibitem{Xiao:2013iha} 
  T.~Xiao, S.~Dobbs, A.~Tomaradze and K.~K.~Seth,
  Phys.\ Lett.\ B {\bf 727}, 366 (2013).

\bibitem{epjc-review} N. Brambilla {\em et al.},
\Journal\EPJC{71}{1534}{2011}.

\bibitem{four-quark} Eric Swanson, Physics {\bf 6}, 69 (2013).

\bibitem{bes3-x3872} M. Ablikim {\em et al.} (BESIII Collaboration),
Phys. Rev. Lett. {\bf 112}, 092001 (2014).

\bibitem{3d2-decay} E. J. Eichten, K. Lane, and C. Quigg,
Phys. Rev. Lett. {\bf 89},162002 (2002); P. Cho and M. B. Wise,
Phys. Rev. D {\bf 51}, 3352 (1995).

\bibitem{E705} L. Antoniazzi {\em et al.} (The E705 Collaboration),
\Journal\PRD{50}{4258}{1994}.

\bibitem{belle-3d2} V. Bhardwaj {\em et al.} (Belle Collaboration),
\Journal\PRL{111}{032001}{2013}.

\bibitem{bepc2} M. Ablikim {\em et al.} (BESIII Collaboration),
Nucl. Instrum. Methods Phys. Res., Sect. A {\bf 614}, 345 (2010).

\bibitem{lum} M. Ablikim {\em et al.} (BESIII Collaboration),
arXiv:1503.03408 [hep-ex] (2015). 

\bibitem{geant4} S. Agostinelli {\em et al.} (GEANT4 Collaboration), Nucl. Instrum.
Methods A {\bf 506}, 250 (2003).


\bibitem{evtgen} D. J. Lange, Nucl. Instrum. Methods A {\bf 462}, 152 (2001).

\bibitem{kkmc} S. Jadach, B. F. L. Ward, and Z. Was,
Comput. Phys. Commun. {\bf 130}, 260 (2000); Phys. Rev. D {\bf
63}, 113009 (2001).

\bibitem{y4360} 
X. L. Wang {\em et al.} (Belle Collaboration), arXiv:1410.7641.

\bibitem{photos} P. Golonka, and Z. Was, Eur. Phys. J. C {\bf 45}, 97
(2006).

\bibitem{bes3-etajpsi} M.~Ablikim {\em et al.} (BESIII Collaboratioin),
Phys. Rev. D {\bf 86}, 071101(R) (2012).

\bibitem{Mppjpsi} In this Letter, $M(\ppjpsi)=M(\pp\LL)-M(\LL)+m(\jpsi)$
is used to partly cancel the mass resolution of the lepton pair. Here
$m(\jpsi)$ and $m(\psip)$ are the nominal masses 
of $\jpsi$ and $\psip$~\cite{pdg}.

\bibitem{Kolmogorov} A. Kolmogorov, G. Inst. Ital. Attuari. {\bf 4}, 83
(1933).

\bibitem{VP} F. Jegerlehner, Z. Phys. C {\bf 32} (1986) 195.

\bibitem{rad} E. A. Kuraev and V. S. Fadin, Yad. Fiz. {\bf 41}, 733
(1985) [Sov. J. Nucl. Phys. {\bf 41}, 466 (1985)].

\bibitem{y4360babar} J. P. Lees {\em et al.} (BaBar Collaboration),
Phys. Rev. D {\bf 89}, 111103 (R) (2014).

\bibitem{trigger} N. Berger {\em et al.}, Chin. Phy. C, {\bf 34 (12)}, 1779-1784
(2010).

\bibitem{barnes} T. Barnes, S. Godfrey and E. S. Swanson,
Phys. Rev. D {\bf 72}, 054026 (2005).

\bibitem{3d2-mass} S. Godfrey and N. Isgur, Phys. Rev. D
{\bf 32}, 189 (1985); W. Kwong, J. Rosner, and C. Quigg, Annu.
Rev. Nucl. Part. Phys. {\bf 37}, 343 (1987); D. Ebert, R. N.
Faustov, and V. O. Galkin, Phys. Rev. D {\bf 67}, 014027 (2003);
E. J. Eichten, K. Lane, and C. Quigg, Phys. Rev. D {\bf 69},
094019 (2004); M. Blank and A. Krassnigg, Phys. Rev. D {\bf 84},
096014 (2011).

\bibitem{ratio} C. F. Qiao, F. Yuan, and K. T. Chao,
Phys. Rev. D {\bf 55}, 4001 (1997).

\end{thebibliography}
\end{document}